\documentclass[aps,pre,a4,floatfix,groupedaddress,showpacs,showkeys,twocolumn]{revtex4-1}  
\usepackage{graphicx,amsmath,amssymb,dsfont,bm}

\begin{document}

\title{Cross--response in correlated financial markets: individual stocks}
\author{Shanshan Wang}
\email{shanshan.wang@uni-due.de}
\affiliation{Fakult\"at f\"ur Physik, Universit\"at Duisburg--Essen, Lotharstra\ss e 1, 47048 Duisburg, Germany}
\author{Rudi Sch\"afer}
\email{rudi.schaefer@uni-due.de}
\affiliation{Fakult\"at f\"ur Physik, Universit\"at Duisburg--Essen, Lotharstra\ss e 1, 47048 Duisburg, Germany}
\author{Thomas Guhr}
\email{thomas.guhr@uni-due.de}
\affiliation{Fakult\"at f\"ur Physik, Universit\"at Duisburg--Essen, Lotharstra\ss e 1, 47048 Duisburg, Germany}

\date{\today}

\begin{abstract}
Previous studies of the stock price response to trades focused on the dynamics of single stocks, \textit{i.e.} they addressed the self--response. We empirically investigate the price response of one stock to the trades of other stocks in a correlated market, \textit{i.e.} the cross--responses. How large is the impact of one stock on others and vice versa?  --- This impact of trades on the price change across stocks appears to be transient instead of permanent as we discuss from the viewpoint of market efficiency. Furthermore, we compare the self--responses on different scales and the self-- and cross--responses on the same scale. We also find that the cross--correlation of the trade signs turns out to be a short--memory process.
\end{abstract}


\maketitle

\section{Introduction} 
\label{section1}

The trading at stock exchanges is organized by the order book whose main purpose is to provide the same information to all market participants. Although often ignored in the model building, it has a large impact on the price dynamics and thus on the stylized facts as well as on more specific features~\cite{Cont2001,Chordia2002,Bouchaud2003,Bouchaud2008,Chakraborti2011,Toth2011,Eisler2012,Schmitt2012,Schmitt2013}. The stock price is determined via a continuous double auction~\cite{Farmer2004}, in which some traders submit market orders for immediate transactions at the best available price, while other traders submit limit orders which specify an acceptable price for the trade. The limit orders are listed in the order book.  Most of them do not immediately lead to trades. The buy limit orders are referred to as bids and the sell limit orders as asks. The best ask and best bid prices are the quotes. Market orders do not appear in the order book. When a market order is executed, the quote can either stay unchanged or the best ask (bid) price can go up (down) in the case of a buy (sell) market order. The prices change persistently as they are affected by the incoming market orders. To profit from the price difference between ask and bid, traders emit limit orders which leads to an anti--persistence of prices. As a result of a detailed balance between persistent and anti--persistent, \textit{i.e.}, between super-- and subdiffusive behavior, the price on an intraday scale moves diffusively like a random walk~\cite{Bouchaud2004}.

In recent years, a high auto--correlation of the order flow was empirically found~\cite{Bouchaud2004,Lillo2004,Lillo2005,Toth2015}. The splitting of orders over longer times introduces long memory of the order flow~\cite{Lillo2005} with remarkable persistence. Buy (sell) orders are often followed by more buy (sell) orders. Furthermore, the relation between trades and price changes has received considerable attention~\cite{Chordia2002,Bouchaud2008,Bouchaud2004,Hausman1992,Kempf1999,Dufour2000,Plerou2002,Rosenow2002,Bouchaud2006,Mike2008}. The Efficient Market Hypothesis (EMH)~\cite{Fama1970} states that all available information is processed and encoded in the current price, which would rule out any (statistical) arbitrage opportunities. While this is in conflict with the very different time scales on which, first, relevant new information arrives and, second, the prices change, the model of Zero Intelligence Trading (ZIT)~\cite{Gode1993} simply assumes randomly acting trader, but also arrives at a memoryless random walk.

In view of the EMH, there are two major approaches to explain the impact of trades on the stock price change. The first approach put forward by Lillo and Farmer (LF)~\cite{Lillo2004}, suggests that the price impact is permanent, but fluctuates with order size. The impact is caused by an asymmetry in liquidity which is induced by the trade. The self--response exhibits a power--law relation between order size and price change~\cite{Lillo2004,Lillo2003,Gabaix2003,Plerou2004}. In the
second approach, Bouchaud, Gefen, Wyart and Potters (BGPW)~\cite{Bouchaud2004} argue that the price impact is transient, but fixed with order size. The fact that the impact decays with time is a result of price mean reversion. Moreover, they identify the relation between order size and price self--response as logarithmic~\cite{Potters2003}. Gerig~\cite{Gerig2008} suggests that the two approaches LF and BGPW are equivalent and can be related by exchanging variables. He also argues that the impact comes from the asymmetric liquidity rather than from the price mean reversion.

There are numerous studies devoted to the price response, but they all focus on one single stock, \textit{i.e.} on the self--responses. Here, we go beyond this and investigate the role of correlations. We carry out a large--scale empirical study of real--time trade data and find a non--vanishing price response across different stocks, \textit{i.e.} for the cross--responses. We shed light on the price impact from trades in different stocks by discussing the efficiency of the financial market. We thereby present a first complete view of the response in the market as a whole and identify several structural characteristics.

The paper is organized as follows. In Sect.~\ref{section2}, we present our data set of stocks, provide some basic definitions, and introduce two possible time scales for the data analysis. We also test the accuracy of the trade sign classification.  In Sect.~\ref{section3}, we show the empirical results, which demonstrate the existence of trade sign cross--correlations and price cross--response, and we also compare two possible definitions of the response functions, in-- or excluding the zero trade signs. In Sect.~\ref{section4}, we present the market response structures and discuss the trade impact on the prices in detail, especially from the viewpoint of market efficiency. In Sect.~\ref{section5}, we compare the self--responses based on the two different time scales, and we also compare the self-- and cross--responses. We give our conclusions in Sect.~\ref{section6}.

\section{Data description and time convention}   
\label{section2}

In Sect.~\ref{sec21}, we present the data set that we use in our analysis. In Sect.~\ref{sec22} we discuss the proper choice of time convention. In Sect.~\ref{sec23}, we test the accuracy of our trade sign classification.

\subsection{Data set}
\label{sec21}

Our study is based on the data from NASDAQ stock market in the year 2008. NASDAQ is a purely electronic stock exchange, whose Trades and Quotes (TAQ) data set contains the time, price and volume. This information is not only given for the trades with all successive transactions, but also for the quotes with all successive best buy and sell limit orders.

To investigate the response across different stocks in Sect.~\ref{section3}, we select six companies from three different economic sectors traded in the NASDAQ stock market in 2008. The stocks we analyzed are listed in Table~\ref{tab21} together with their acronyms and the corresponding economic sectors.
\begin{table} [b]
\linespread{0.5} 
\begin{center}
\caption{Company information} 
\begin{tabular}{lll}
\hline
\hline
Company				&Symbol	&Sector	\\
\hline
Apple Inc.           		&AAPL   	&Information technology\\
Microsoft Corp.      		&MSFT   	&Information technology\\
Goldman Sachs Group	&GS     	&Financials\\
JPMorgan Chase       	&JPM    	&Financials\\
Exxon Mobil Corp.    		&XOM    	&Energy\\
Chevron Corp.        		&CVX     	&Energy\\
\hline
\hline
\label{tab21}
\end{tabular}
\end{center}
\vspace*{-0.6cm} 
\end{table}

When studying the market response in Sect.~\ref{section4}, we select the first ten stocks with the largest average market capitalization in each economic sector of the S$\&$P 500 index in 2008, except for the telecommunications services where only nine stocks were available in that year. We recall that the market capitalization is the trade price multiplied with the traded volume, and the average is performed over every trade during the year 2008. The selected 99 stocks are listed in App.~\ref{appA}.

We only consider the common trading days in which the trading of stocks $i$ and $j$ took place, because the trades of one stock $i$ in one day would not impact the intraday price of another stock $j$ without any trade in that day, and \textit{vice versa}.

\begin{figure*}[htbp]
  \begin{center}
    \includegraphics[width=0.85\textwidth]{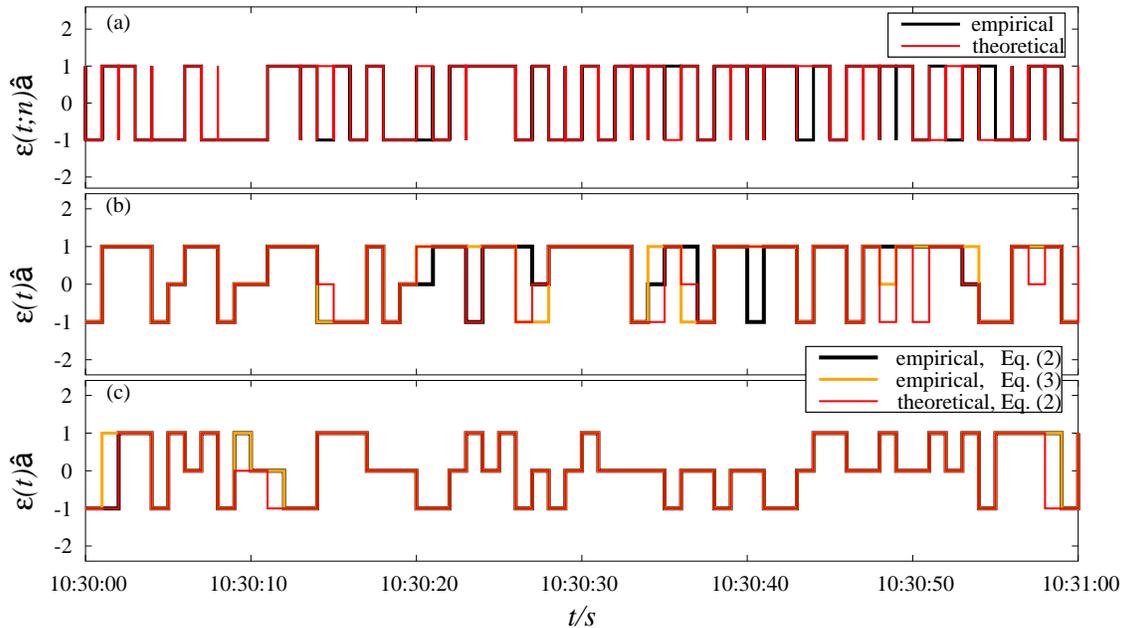}
    \caption{Comparison of different classifications for empirical and theoretical trade signs versus the physical time for AAPL during one minute. (a) Sign comparison of every trade for AAPL on January 7th, 2008. (b) Sign comparisons of every second for AAPL on January 7th, 2008 with the worst accuracy difference of 5$\%$ out of six samples. (c) The sign comparisons of every second for AAPL on June 2nd, 2008. It shows a typical accuracy difference of 2$\%$.}
   \label{fig21}
   \end{center}
   \vspace*{-0.05cm}
\end{figure*}

\subsection{Physical versus trading time}
\label{sec22}

While studies on the self--responses employ trading time as time axis, this is not useful when studying the response across different stocks, because each stock has its own trading time. Hence, to study the cross--responses, we better use the real, physical time. We project the data set to a discrete time axis. The quote data and the trade data of each stock are in two separate files with a time--stamp accuracy of one second. However, more than one quote or trade may be recorded in the same second. Due to the one--second accuracy of the time--stamps, it is not possible to match each trade with the directly preceding quote. Hence, we cannot determine the trade sign by comparing the traded price and the preceding midpoint price. This latter definition of the trade sign was employed by Lee and Ready~\cite{Lee1991}. Instead, we here define the trade signs similarly to the tick rule of Holthausen, Leftwich and Mayers~\cite{Holthausen1987}. They define the trade as buyer--initiated (seller--initiated) if the trade is carried out at a price above (below) the prior price. Zero tick trades are not classified in general. The tick rule has an accuracy of $52.8\%$~\cite{Holthausen1987}. For our study, we further develop this method: as our data has a one--second accuracy in time, we consider the consecutive time intervals of length one second. Let $t$ label such an interval and let $N(t)$ be the number of trades in that interval. The individual trades carried out in this interval are numbered $n=1,\ldots,N(t)$ and the corresponding prices are $S(t;n)$. We define the sign of the price change between consecutive trades as
\begin{eqnarray}       
\varepsilon(t;n)=\left\{                  
\begin{array}{lll}    
\mathrm{sgn}\bigl(S(t;n)&-&S(t;n-1)\bigr)   \ ,  \\
                                      &&~\mbox{if}~~S(t;n)\neq S(t;n-1) , \\    
            \varepsilon(t;n-1) \ ,&&~\mbox{otherwise}.
\end{array}           
\right.  
\label{eq21}            
\end{eqnarray}
If two consecutive trades of the same trading direction together did not exhaust all the available volume at the best quote, the prices of both trades would be the same. Thus, we set the trade sign equal to the previous trade sign in this case. If there is more than one trade in the interval denoted $t$, we average the corresponding trade signs,
\begin{eqnarray}       
\varepsilon(t)=\left\{                  
\begin{array}{lll}    
\mathrm{sgn}
\left(\sum\limits_{n=1}^{N(t)}\varepsilon(t;n)\right) & \ , \quad & \mbox{if} \quad N(t)>0 \ , \\    
                                                0 & \ , \quad & \mbox{if} \quad N(t)= 0 \ ,
\end{array}           
\right. 
\label{eq22}              
\end{eqnarray}     
which formally also includes the case $N(t)=1$. Consequently $\varepsilon(t)=+1$ implies that the majority of trades in second $t$ was triggered by a market order to buy, and a value $\varepsilon(t)=-1$ indicates a majority of sell market orders. We have $\varepsilon(t)=0$, whenever trading did not take place in the time interval $t$ or if there was a balance of buy and sell market orders, \textit{i.e.} if the argument of the sign function vanishes, $\mbox{sgn}(0)=0$. In order to avoid overnight effects and any artifacts at the opening and closing of the market, we consider only trades of the same day from 9:40:00 to 15:50:00 New York local time.

\subsection{Accuracy of the trade sign classification}
\label{sec23}

\begin{table*} [htbp]
\begin{center}
\caption{Accuracy of trade sign classification} 
\begin{tabular}{l@{\hskip 0in}r@{\hskip 0.1in}r@{\hskip 0.1in}r@{\hskip 0.1in}r@{\hskip 0.1in}r@{\hskip 0.1in}r@{\hskip 0.2in}c} 
\hline
\hline
Stock 			&AAPL 		&AAPL 		&GS 		&GS 		&XOM 		&XOM 		& six samples\\
Date 			&20080107	&20080602	&20081007	&20081210	&20080211	&20080804 	& (average)  \\
\hline
\multicolumn{8}{l}{For consecutive trades \footnotemark[1]}\\
Number of identified limit orders 				&745020 	&407843 	&150532 	&199224 	&544451 	&596882 	&\\
Number of identified trades 		&120287 	&52691 	&19942 	&17902 	&38455 	&59580 	&\\
Number of matches 		 		&103635 	&47748 	&16668 	&15454 	&30478 	&49921 	&\\
Accuracy of the classification			&0.86 	&0.91 	&0.84 	&0.86 	&0.79 	&0.84 	&0.85\\
\\
\multicolumn{8}{l}{For trades with the stamp of one second  \footnotemark[1] \footnotemark[2]}\\
Total number of identified trade signs 					&17115 	&12180 	&8283 	&6853 	&8782 	&9590 	&\\
Number of matches for Eq.~\eqref{eq22} 	&13956 	&10636 	&6801 	&5784 	&6516 	&7777 	&\\
Accuracy of the classification for Eq.~\eqref{eq22} 		&0.82 	&0.87 	&0.82 	&0.84 	&0.74 	&0.81 	&0.82\\
Number of matches for Eq.~\eqref{eq23} 	&13256 	&10302 	&6715 	&5690 	&6446 	&7603 	&\\
Accuracy of the classification for Eq.~\eqref{eq23} 		&0.77 	&0.85 	&0.81 	&0.83 	&0.73 	&0.79 	&0.80 \\
\\
\multicolumn{8}{l}{Total number of $\varepsilon(t)=0$ found empirically \footnotemark[1]}\\
Using Eq.~\eqref{eq22}	&6000 	&10515 	&14218 	&15512 	&13719 	&12866	& 12138 \\
Using Eq.~\eqref{eq23}	&5343 	&10186 	&14051 	&15426 	&13571 	&12731	& 11885 \\
\hline
\hline
\footnotetext[1]{The trading time in each day is set to 9:40:00 to 15:50:00 New York local time (total 22200 seconds).}\\
\footnotetext[2]{The case that $\varepsilon(t)=0$ occurs simultaneously in three kinds of trade signs is excluded.}
 \label{tab22}
\end{tabular}
\end{center}
\vspace*{-1.2cm}
\end{table*}

We test the accuracy of the trade sign classification with the intraday data of AAPL, GS and XOM from NASDAQ stock market by analyzing the TotalView--ITCH data set \cite{Web}, which provides the information including buy or sell type, order price, and order volume for each limit order with a unique order ID on the scale of milliseconds. With this data set, we can identify the trade directions (buy or sell) of the market orders, which show opposite trade directions to the limit orders executed simultaneously. For example, an executed sell limit order corresponds to a buyer--initiated market order. The executed prices and volumes of the market orders can be obtained as well according to the information of the executed limited orders. Here, we regard an execution of one limit order as a transaction accompanied with a market order in chronological order. Therefore, the trade signs of the market orders inferred from the types (buy or sell) of the executed limited orders are referred to as the empirical trade signs, while the trade signs achieved by comparing the prices between consecutive trades, as in Eq.~\eqref{eq21}, are referred to as the theoretical trade signs. Although the TotalView--ITCH data set indirectly provides the empirical trade signs, it does not give the information of the best quote and the trade as conveniently as the TAQ data set. Thus, we only use the TotalView--ITCH data set for testing the accuracy of the trade sign classification.

In the TotalView--ITCH data set, the executed orders can be classified as non--displayed orders and displayed limit orders in the order book. The executed non--displayed orders correspond to the hidden trades that we cannot use in our study for testing the accuracy of the trade sign classification. The executed limit orders correspond to the trades identified as buyer-- or seller--initiated. We refer to these trades as to the identified trades. With these identified trades, we compare the theoretical trade signs $\varepsilon(t;n)$ with the empirical ones to test the accuracy. For a given trade, we count the number of matches, \textit{i.e.}  the number of cases in which the empirical and theoretical signs are the same. Thus, the accuracy of the trade sign classification for a given trade is defined as the number of matches divided by the total number of identified trades.

We randomly select two trading days for each stock in 2008. Due to high--frequency trading, there are more than 10000 trades for each stock executed in each trading day. As shown in Table~\ref{tab22}, the average accuracy of sign classification is equal to $85\%$ for six tested samples. Fig.~\ref{fig21} (a) compares the two kinds of trade signs for AAPL during one minute at Jan. 7th, 2008. The theoretical trade signs nicely match with the empirical ones.

Furthermore, we test the trade sign $\varepsilon(t)$ of every second, defined as the sign of the number imbalance of trades in one second as in Eq.~\eqref{eq22}. We use this definition to calculate our theoretical trade sign in every second. For the empirical trade sign in every second, we evaluate the aggregated trade sign in two ways. We once more use Eq.~\eqref{eq22} and also employ the sign of the volume imbalance of trades~\cite{Plerou2002,Gabaix2003,Plerou2004}, as given by
\begin{eqnarray}       
\varepsilon(t)=\left\{                  
\begin{array}{lll}    
\mathrm{sgn}
\left(\sum\limits_{n=1}^{N(t)}\varepsilon(t;n)v(t;n)\right) & ,  & \mbox{if} \quad N(t)>0 \ , \\    
                                                0 & ,  & \mbox{if} \quad N(t)= 0 \ ,
\end{array}           
\right. 
\label{eq23}              
\end{eqnarray}
where $v(t;n)$ is the trading volume of $n$--th trade in second $t$. If the volume imbalance, \textit{i.e.} the argument of the sign function, is a positive (negative) value, then $\varepsilon(t)=+1$ ($-1$) implies buyer (seller)--initiated market orders in $t$. If it is equal to zero, $\varepsilon(t)=0$ indicates a volume balance of buy and sell market orders. Moreover, $\varepsilon(t)=0$ also means that there was not any trade in $t$.

The total trading time in each trading day is 22200 seconds. In case that the accuracy of the sign classification is influenced by an excess of zero trade signs, the trading times in which $\varepsilon(t)=0$ simultaneously occurs in the three kinds of trade sign classification are excluded. Thus, the remanent trading time is used to test the accuracy of the classification of the trade signs $\varepsilon(t)$. We refer to this remanent trading time as to the identified trading time. For each tested stock, there are more than 6000 seconds of identified trading time as shown in Table~\ref{tab22}. That means there are more than 6000 identified trade signs for each tested stock, since every second has a trade sign. For each second, we compare the theoretical trade sign with the empirical one. Again, we count the number of matches. Thus, the accuracy of the trade sign classification for every second is defined as the number of matches divided by the total number of the identified trade signs.

When the theoretical trade signs are compared with the empirical trade signs aggregated according to Eq.~\eqref{eq22}, the average accuracy of six tested samples in Table~\ref{tab22} is $82\%$. When the theoretical trade signs are compared with the empirical ones aggregated according to Eq.~\eqref{eq23}, the average accuracy is $80\%$. These two scenarios have an accuracy difference of $2\%$ only. That means the theoretical trade signs defined by Eq.~\eqref{eq22} have a high number of matches with the empirical trade signs aggregated either according to Eq.~\eqref{eq22} or to Eq.~\eqref{eq23}. Moreover, the different ways of aggregating trade signs do not strongly influence the accuracy of the trade sign classification. Figs.~\ref{fig21} (b) and (c) show the comparisons of three kinds of trade signs for AAPL on the scale of one second. Compared to the other five samples, AAPL at January 7th, 2008 in (b) shows the worst accuracy difference of $5\%$. However, the matches for three kinds of trade signs still can be found most of the time. As a typical example, AAPL at June 2nd, 2008 in (c) shows the general case with the accuracy difference of 2$\%$, where the theoretical trade signs match with the two kinds of aggregated empirical trade signs rather well.

\begin{table*}[htbp]
\linespread{0.5} 
  \caption{Fit parameters and normalized $\chi_{ij}^2$ for the trade sign cross--correlators.} 
\begin{center}
\begin{tabular}{c@{\hskip 0.2in}l@{\hskip 0.2in}l@{\hskip 0.2in}c@{\hskip 0.1in}cc@{\hskip 0.2in}c@{\hskip 0.1in}cc@{\hskip 0.2in}c@{\hskip 0.1in}cc@{\hskip 0.2in}c@{\hskip 0.1in}c} 
\hline
\hline
sign	&stock $i$ ~&stock $j$ & \multicolumn{2}{c}{$\vartheta_{ij}$} && \multicolumn{2}{c}{$\tau_{ij}^{(0)}$~[ s ]}&&\multicolumn{2}{c}{$\gamma_{ij}$ }& & \multicolumn{2}{c}{$\chi_{ij}^2$~($\times10^{-6}$)} \\
\cline{4-5}\cline{7-8}\cline{10-11}\cline{13-14}
correlators&	&			&inc. 0	&exc. 0	&&inc. 0	&exc. 0	&&inc. 0	&exc. 0	&&inc. 0	&exc. 0	\\
\hline
	&AAPL	&MSFT 		&0.46	&0.05	&&0.05	&3.46	&&1.00	&1.35	&&0.23 	&1.52	\\
	&MSFT	&AAPL		&0.04	&0.07	&& 2.34	&2.34	&&1.15 	&1.15	&&0.10	&0.27	\\
	&XOM	& CVX		&0.61	&0.67	&&0.06	&0.21	&&1.04 	&1.16	&&0.07	&0.52	\\
cross&GS		&JPM		&0.45	&0.48	&&0.07	&0.13	&&1.00	&1.00	&&0.04	&0.18	\\
	&AAPL	&GS			&0.46	&0.28	&&0.03	&0.14	&&1.00 	&0.91	&&0.11	&0.99	\\
	&GS		&AAPL		&0.49 	&0.49	&&0.06	&0.10	&&1.00	&1.00	&&0.05 	&0.13	\\
	&GS		&XOM		&0.61	&0.73	&&0.04	&0.08	&&1.04	&1.10	&&0.04	&0.20	\\
	&XOM	&AAPL		&0.76	&0.29	&&0.05	&0.34	&&1.09	&1.42	&&0.12	&0.18	\\
\hline
	&AAPL	&AAPL 		&0.60	&0.96	&&0.21	&0.21	&&1.27	&1.27	&&0.18 	&0.50	\\
self	&GS		&GS 		&0.54	&0.71	&&0.12	&0.25	&&1.17	&1.18	&&0.04 	&0.44	\\
	&XOM	&XOM 		&0.54	&0.89	&&0.17	&0.23	&&1.12	&1.14	&&0.09 	&0.49	\\
\hline
\hline
\end{tabular}
\end{center}
\vspace*{-0.5cm} 
\label{tab31}
\end{table*}

\section{Cross--responses for pairs of stocks} 
\label{section3}

To study the mutual dependences between stocks, we consider pairs of two different stocks. We introduce the cross--response function as well as a trade sign cross--correlator for such stock pairs in Sect.~\ref{sec31} and Sect.~\ref{sec32}, respectively. It turns out that the in-- or exclusion of the trade signs $\varepsilon(t)=0$ make
a difference. We compare these two possible definitions in Sect.~\ref{sec33}.  In the sequel, all quantities referring to a particular stock carry its index $i$, quantities referring to a pair carry two such indices. We consider eight pairs of stocks listed in Table~\ref{tab21}, four within the same economic sector, four across different economic sectors.

\subsection{Cross--response functions} 
\label{sec31}

\begin{figure*}[htbp]
  \begin{center}
   \includegraphics[width=0.95\textwidth]{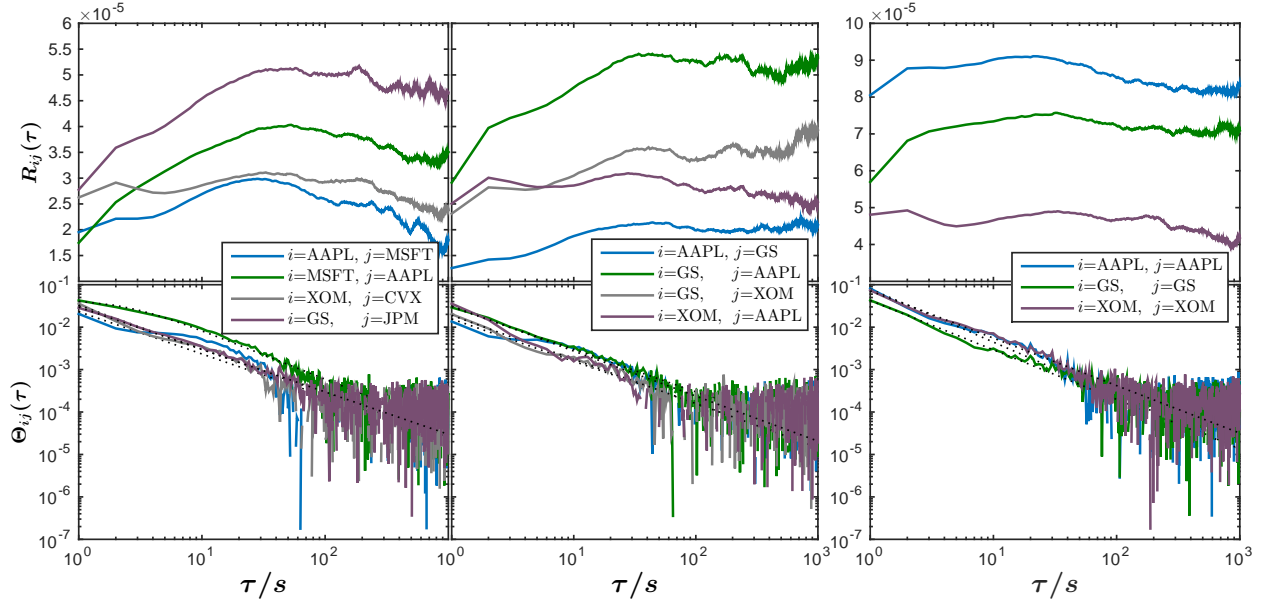}
  \end{center}
  \vspace*{-0.5cm}
  \caption{Cross--response functions $R_{ij}(\tau)$ including $\varepsilon_j(t)=0$ in 2008 versus time lag $\tau$ on a logarithmic scale (top panels).  Corresponding trade sign cross--correlators $\Theta_{ij}(\tau)$ for different stock pairs on a doubly logarithmic scale, fit as dotted lines (bottom panels). The stock pairs in the first column of panels are from the same economic sectors, and in second column of panels are from the different economic sectors. The third column of panels are the self--responses and sign self--correlators to be compared with cross--responses and sign cross--correlators.}
\label{fig311}
\end{figure*}

\begin{figure*}[htbp]
  \begin{center}
   \includegraphics[width=0.95\textwidth]{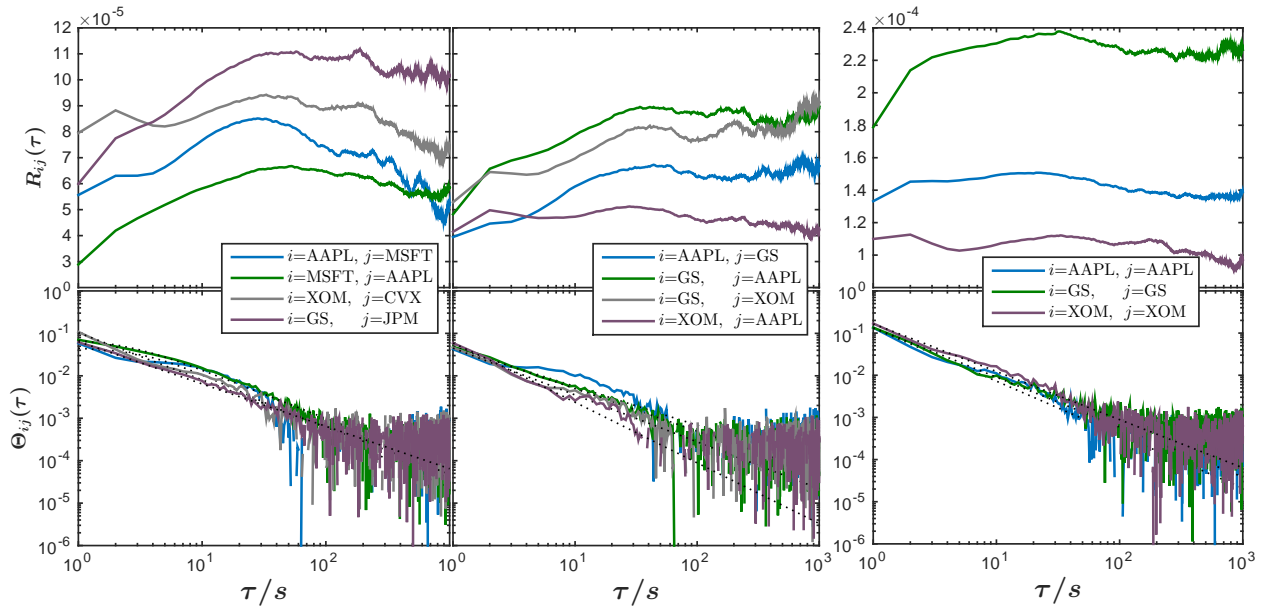}
  \end{center}
  \vspace*{-0.5cm}
  \caption{Cross--response functions $R_{ij}(\tau)$ excluding $\varepsilon_j(t)=0$ in 2008 versus time lag $\tau$ on a logarithmic scale (top panels).  Corresponding trade sign cross--correlators $\Theta_{ij}(\tau)$ for different stock pairs on a doubly logarithmic scale, fit as black dotted lines (bottom panels). The stock pairs in the first column of panels are from the same economic sectors, and in second column of panels are from the different economic sectors. The third column of panels are the self--responses and sign self--correlators to be compared with cross--responses and sign cross--correlators.}
\label{fig312}
\end{figure*}

To measure how a buy or sell order of stock with index $j$ at time $t$ influences the prices of the stock $i$ at a later time $t+\tau$, we introduce the cross--response function. We employ the logarithmic price differences or log--returns for stock $i$ and time lag $\tau$, defined via the midpoint prices $m_i(t)$,
\begin{equation}
r_i(t,\tau) \ = \ \log m_i(t+\tau) -\log m_i(t) \ = \ \log\frac{m_i(t+\tau)}{m_i(t)}
\label{eq31}
\end{equation}
at a given time $t$, keeping in mind the one--second accuracy.  To acquire statistical significance, the cross--response function is the time average
\begin{equation}
R_{ij}(\tau) \ = \ \Bigl\langle r_i(t,\tau)\varepsilon_j(t)\Bigr\rangle _t \ 
\label{eq32}
\end{equation}
of the product of time--lagged returns and trade signs for stocks $i$ and $j$, respectively. Two definitions are possible and meaningful: one can in-- or exclude the trade signs $\varepsilon_j(t)=0$. This does make a difference as it affects the normalization: The total number of events that determines the normalization constant for the average is larger when including the events with $\varepsilon_j(t)=0$, although they yield contributions $r_i(t,\tau)\varepsilon_j(t)=0$. We further discuss this issue in Sect.~\ref{sec33}. Here, we begin with presenting our empirical results for different stock pairs $(i,j)$ in Figs.~\ref{fig311} and \ref{fig312} versus the time lag, for in-- and excluding trade signs $\varepsilon_j(t)=0$, respectively. In all cases, an increase to a maximum is followed by a decrease, \textit{i.e.} the trend in the cross--response is eventually reversed.

This trend does not depend on whether or not the pairs are in the same economic sector or extend over two sectors. The stocks face similar systematic risks, leading to stronger cross--response in the same sector than across different sectors. However, strong cross--responses for the stock pairs from different sectors also exist, \textit{e.g.} for (GS, AAPL). Apart from reasons specific for the stock pair considered, this might also be related to how investors assemble their portfolios. To disperse the investment risks, the portfolios often comprise stocks from different sectors since they are exposed to different economic risks and are less correlated than stocks within the same sector. When investors buy or sell the stocks in their portfolios gradually, it may produce cross--responses and sign cross--correlations in different stocks. We measure the strength of the sign cross--correlation in Sect.~\ref{sec32}.

The price reversion is also independent of in-- or excluding zero trade signs $\varepsilon_j(t)=0$. The cross--response including $\varepsilon_j(t)=0$ measures the remanent price impact of market orders of stock $j$ taking into account the cases in which trading did not occur, while the cross--response excluding $\varepsilon_j(t)=0$ purely measures the price impact of market orders ignoring the lack of trading. The difference of both cross--response functions is mainly in the overall amplitude, but the general trends for each stock pair are quite similar. More details will be given in Sect.~\ref{sec33}.

As we quantify the price impact to every second, the non--zero value of $R_{ij}(1)$ represents the one--second impact of trades between different stocks. It is rather different from the instantaneous impact in single stocks, as the trades in our study do not consume the volumes in order book of impacted stocks to shift the price directly. The price impact between stocks may come from other mechanisms, such as trading information transmission, which affects the limit orders of impacted stocks placed or cancelled, or even from market orders executed to move the price indirectly. This is possible due to high frequency trading, which leads to trades executed at the level of milliseconds. Thus, the time interval of one second is sufficient to identify trading information transmission or other mechanisms.

It is worth mentioning that Figs.~\ref{fig311} and \ref{fig312} show an increase of the cross--response after decreasing back at large time lag $\tau$ close to 1000 s. We attribute this to the response noise, quantified in App.~\ref{appC}.

\subsection{Trade sign cross--correlator}
\label{sec32}

The existence of sign self--correlations is the main reason that causes the self--response~\cite{Bouchaud2004}. For pairs of stocks, we introduce the trade sign cross--correlator
\begin{equation}
\Theta_{ij}(\tau) \ = \ \Bigl\langle \varepsilon_i(t+\tau)\varepsilon_j(t) \Bigr\rangle _t
\label{eq33}
\end{equation}
as a function of the time lag $\tau$. To study how the in-- or exclusion of $\varepsilon_j(t)=0$ impacts cross--correlations of trade signs, we also distinguish the two possible definitions. However, it turns out that the differences are negligible. The sum of the product of trade signs between stocks is not changed. What changes is the
total number of trades, which enlarges or shrinks the average value. As seen in Figs.~\ref{fig311} and \ref{fig312}, the inclusion of $\varepsilon_j(t)=0$ actually decreases the sign cross--correlation instead of increasing.

As we demonstrate in Figs.~\ref{fig311} and \ref{fig312}, there is a non--zero correlation across stocks. It turns out that the empirical result can be fitted well by the power law
\begin{equation}
\Theta_{ij}(\tau)=\frac{\vartheta_{ij}}{\left(1+(\tau/\tau_{ij}^{(0)})^2\right)^{\gamma_{ij}/2}} \ .
\label{eq34}
\end{equation}
To estimate the error, we use the normalized $\chi_{ij}^2$, see App.~\ref{appB}. The parameters for the best fit as well as the $\chi_{ij}^2$ values for the analyzed eight stock pairs are listed in Table~\ref{tab31}. In contrast to the sign self--correlation on the trading time scale~\cite{Bouchaud2004,Lillo2004}, most of the stock pair cross--correlations on the physical scale exhibit short memory with exponents $\gamma_{ij}\geq1$ rather than long memory. The latter usually is defined as corresponding to exponents smaller than unity~\cite{Beran1994}. This indicates that the price change of one stock responding to the trades of another stock only persists for shorter times, and the cross--response reverses at relatively small time lags $\tau$. When comparing the sign cross--correlator including $\varepsilon_j(t)=0$ with the one excluding $\varepsilon_j(t)=0$, it is instructive to look at the parameter $\tau_{ij}^{(0)}$, which measures the decay period of sign cross--correlation. For most of the
stock pairs, the sign cross--correlation excluding $\varepsilon_j(t)=0$ shows a longer decay period than that including $\varepsilon_j(t)=0$. This illustrates that the lack of trading or the balance of buy and sell market orders accelerates the decay of the sign cross--correlations.

In addition, we notice the large fluctuations of the trade sign cross--correlator at larger lags $\tau$. They are partly due to the decrease of the response signal, but also to the limited statistics. The larger the time lag $\tau$, the larger is the overlap of the lag $\tau$ for different times $t$. When averaging the sign cross--correlation over every second $t$ for large $\tau$, the result has poor statistics.

\subsection{Including or excluding zero trade signs} 
\label{sec33}

\begin{figure}[b]
  \begin{center}
    \includegraphics[width=0.5\textwidth]{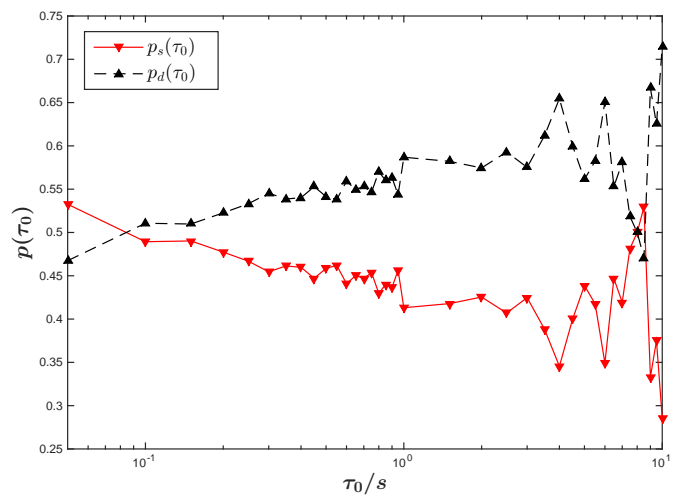}
  \end{center}
  \vspace*{-0.3cm} 
  \caption{Probability densities $p_{s}(\tau_0)$ and $p_{d}(\tau_0)$ for the change of trade sign versus the time $\tau_0$ without trading. The intraday data used stem from five successive trading days of AAPL in 2008. The strong fluctuations at large $\tau_0$ are due to the limited statistics.}
 \label{fig32}
\end{figure}

\begin{figure*}[htbp]
  \begin{center}
  \begin{minipage}[c]{0.7\textwidth}
    \includegraphics[width=1\textwidth]{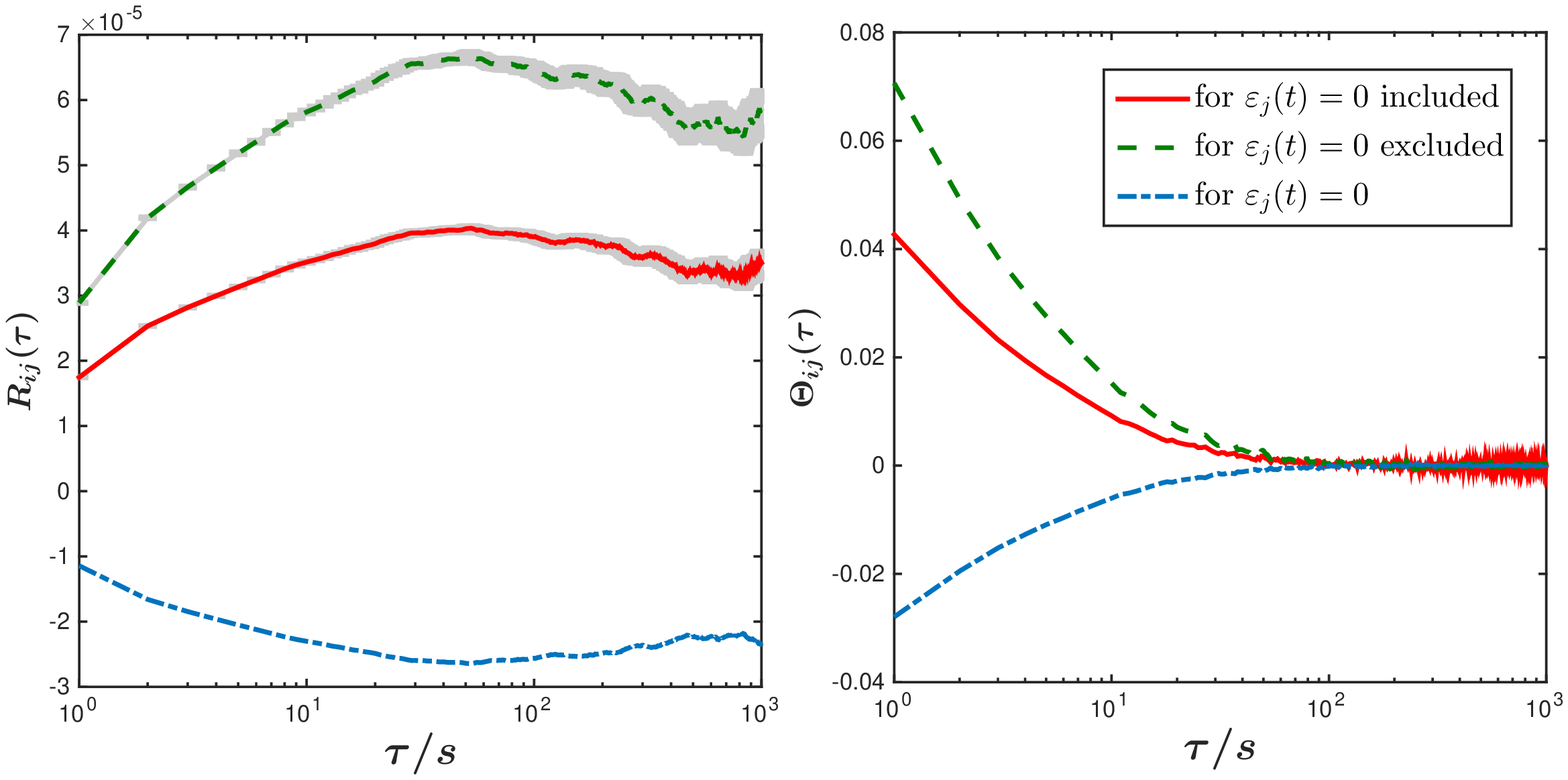}
   \end{minipage}
   \begin{minipage}[c]{0.25\textwidth}
     \caption{The three cross--responses (left) and sign cross--correlators (right) of stock pair (MSFT, AAPL) versus the physical time lag $\tau$: in-- and excluding $\varepsilon(t)=0$ as well as only for $\varepsilon(t)=0$. The shaded regions indicate the standard error bars.}
 \label{fig33}
   \end{minipage} 
  \end{center}
  \vspace*{-0.5cm}
\end{figure*}

Previous studies~\cite{Bouchaud2004,Lillo2004,Toth2015,Gerig2008} focus on the self--response or sign self--correlation. The time used is event, \textit{i.e.} trade based rather than physical. Thus, only buy or sell market orders are considered. In contrast, our study uses the physical time scale meaning that the price impact is quantified to every second, no matter if there was a trade or not. This is necessary or even inevitable as we study cross--responses for trades of different stocks which never are synchronous. For the self--response, one trade can cause the change of the volumes in the order book, leading to further price changes. However, for the cross--response, one trade from a different stock cannot change the price of the impacted stock directly by consuming the volumes of the limit orders in the order book. The necessity to use the physical time forces us to deal with the case $\varepsilon(t)=0$ if a trade does not occur in a given second $t$. The corresponding information can be found in Table~\ref{tab22}, where, on average, more than half of the total physical time for each trading day features zero trade signs.

There are two alternative choices for the sign at $t$ without any trade. One considers $\varepsilon(t)=0$, such that the price at $t$ is not affected. After averaging the response over the entire time, the influence of lack of trades is included in the response function. If the average is restricted to the time in which trading took place, the influence of the lack of trading is obviously excluded. In our opinion, the fact that there was a trade or not cannot be treated as negligible feature of the order book. It contains important information on the trading activity, reflecting that the traders have not traded for some time for whatever economic or other reason. We thus believe that to only consider the impact from successive trades no matter how long the time without trading is between them on the physical time scale might introduce a misleading bias. Many values of $\varepsilon(t)=0$ can be found in each trading day, see Table~\ref{tab22}.

Alternatively, we can simply keep the trade sign until the next trade occurs. This means that, if a trade did not take place, the last trade continues to have an unaltered impact until the next trade is executed. The corresponding picture is reminiscent of Fig.~\ref{fig21} (a). This alternative choice implies that the trade sign which is fixed in the above way at second $t$ is independent of the length of the time $\tau_0$ without trading. Empirically, however, the trade sign is more likely to reverse after a time $\tau_0$ without trading. We demonstrate this for AAPL in Fig.~\ref{fig32} which displays the probability densities $p_{s}(\tau_0)$ and $p_{d}(\tau_0)$ for finding the same or different trade signs, respectively, after $\tau_0$. By definition, we have $p_{s}(\tau_0)+p_{d}(\tau_0)=1$.  We notice that, as the time without trading evolves, the probability density for finding the same sign decreases, while the probability for a sign reversal increases. Thus, the longer the time without trading, the weaker is the impact of the last trade. Obviously, the second choice introduces a bias. Hence, we stick to the first choice, because the response including $\varepsilon(t)=0$ properly relates to the decaying impact, while the response excluding $\varepsilon(t)=0$ yields results similar to using trading time instead of physical time.

Furthermore, we emphasize that $\varepsilon(t)=0$ can have two reasons: either lack of trading or a balance of buy and sell market orders in $t$. In the latter case, although the trade directions of buy and sell market orders in one second annihilate each other, we cannot say whether or not the trading itself, regardless of the direction, causes a cross--response of another stock. If so, does it weaken or strengthen the total cross--response? --- To be more quantitative, we introduce the cross--responses
$R_{ij}^\textrm{(inc. $0$)}(\tau)$ and
$R_{ij}^\textrm{(exc. $0$)}(\tau)$ in-- and excluding $\varepsilon(t)=0$, respectively. Here, we do not distinguish the two reasons for $\varepsilon(t)=0$. Hence, in a formal manner, the cross--response for $\varepsilon(t)=0$ can be quantified as the difference of these two kinds of cross--responses,
\begin{equation}
R_{ij}^\textrm{(only $0$)}(\tau)=R_{ij}^\textrm{(inc. $0$)}(\tau)-R_{ij}^\textrm{(exc. $0$)}(\tau).
\label{eq35}
\end{equation}
Similarly, the sign cross--correlator for $\varepsilon(t)=0$ is given by
\begin{equation}
\Theta_{ij}^\textrm{(only $0$)}(\tau)=\Theta_{ij}^\textrm{(inc. $0$)}(\tau)-\Theta_{ij}^\textrm{(exc. $0$)}(\tau).
\label{eq36}
\end{equation}
From this formal viewpoint, the cross--response for $\varepsilon(t)=0$, \textit{i.e.} the price change conditioned on not trading or a balance of buy and sell market orders has a non--zero value.  As shown in Fig.~\ref{fig33}, both the cross--response and sign cross--correlator for $\varepsilon(t)=0$ have negative values. In other words, the existence of the events $\varepsilon(t)=0$ weakens the cross--response that is purely due to trades. Eliminating the influence of $\varepsilon(t)=0$ enlarges the impact of trades on the price change. In this sense, the cross--response including $\varepsilon(t)=0$ is a conservative estimation of the price impact. Moreover, the price change conditioned on $\varepsilon(t)=0$ shows a similar trend as the price change conditioned on trades, \textit{i.e.}, it reverses at large lags.  This discussion strongly corroborates our procedure of analysis by including $\varepsilon(t)=0$.

In the sequel, to avoid cumbersome phrases, we will use the term lack of trading for both reasons that give $\varepsilon(t)=0$, \textit{i.e.}, including the balance of buy and sell market orders. We will always show the responses in-- and excluding zero trade signs for comparison.

\section{Market response} 
\label{section4}

We explore the market response structures by normalized response matrices with and without zero trade signs in Sect.~\ref{sec41}. Studying the market as a whole, we discuss the market impact of trades in Sect.~\ref{sec42} before we turn to the transient impact from the perspective of market efficiency in Sect.~\ref{sec43}.

\subsection{Market response structure}
\label{sec41}

The cross--response functions and the trade sign cross--correlators we considered up to now give us a kind of microscopic information for stock pairs. It is equally important to investigate how the trading of individual stocks influences the market as a whole. In a first step, we tackle this question by introducing the market response as the matrix $\rho(\tau)$ whose entries are the normalized response functions at a given time lag,
\begin{equation}
\rho_{ij}(\tau) \ = \ \frac{R_{ij}(\tau)}{\textrm{max\,}(|R_{ij}(\tau)|)} \ ,
\label{eq41}
\end{equation}
where the denominator is the maximum over all stock pairs $(i, j)$ for fixed $\tau$. The diagonal elements are the self--responses, and the off--diagonal elements are the cross--responses. The matrix $\rho(\tau)$ is reminiscent of, but should not be mixed up with a matrix of Pearson correlation coefficients. Importantly, the matrix of the market response is not symmetric, $\rho_{ij}(\tau)\neq \rho_{ji}(\tau)$, as two different quantities, the returns and the trade signs, enter the definition Eq.~\eqref{eq32}. Furthermore, the market response reveals information about the time evolution.

\begin{figure*}[htbp]
  \begin{center}
    \includegraphics[width=0.47\textwidth]{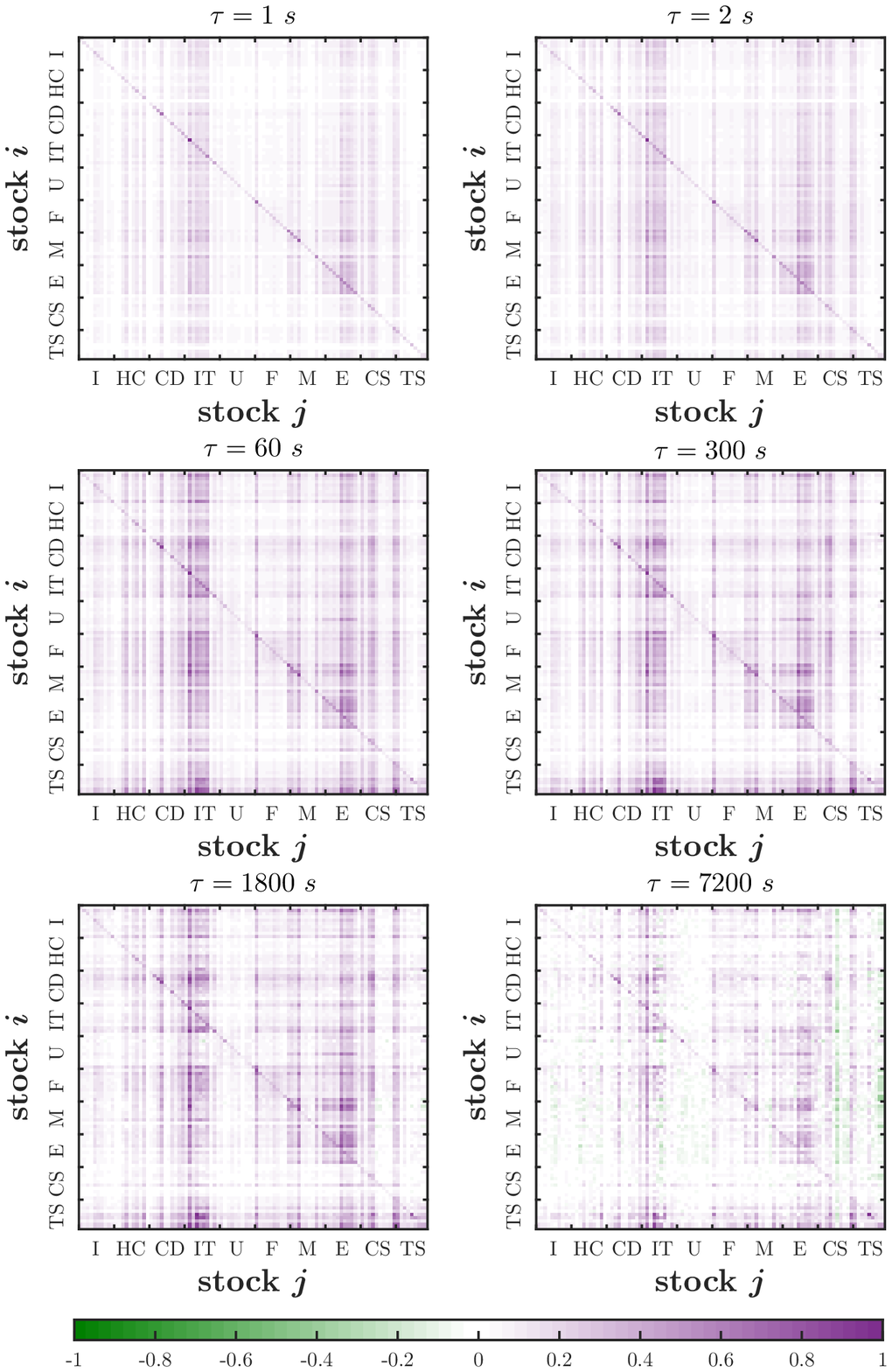}\qquad
    \includegraphics[width=0.47\textwidth]{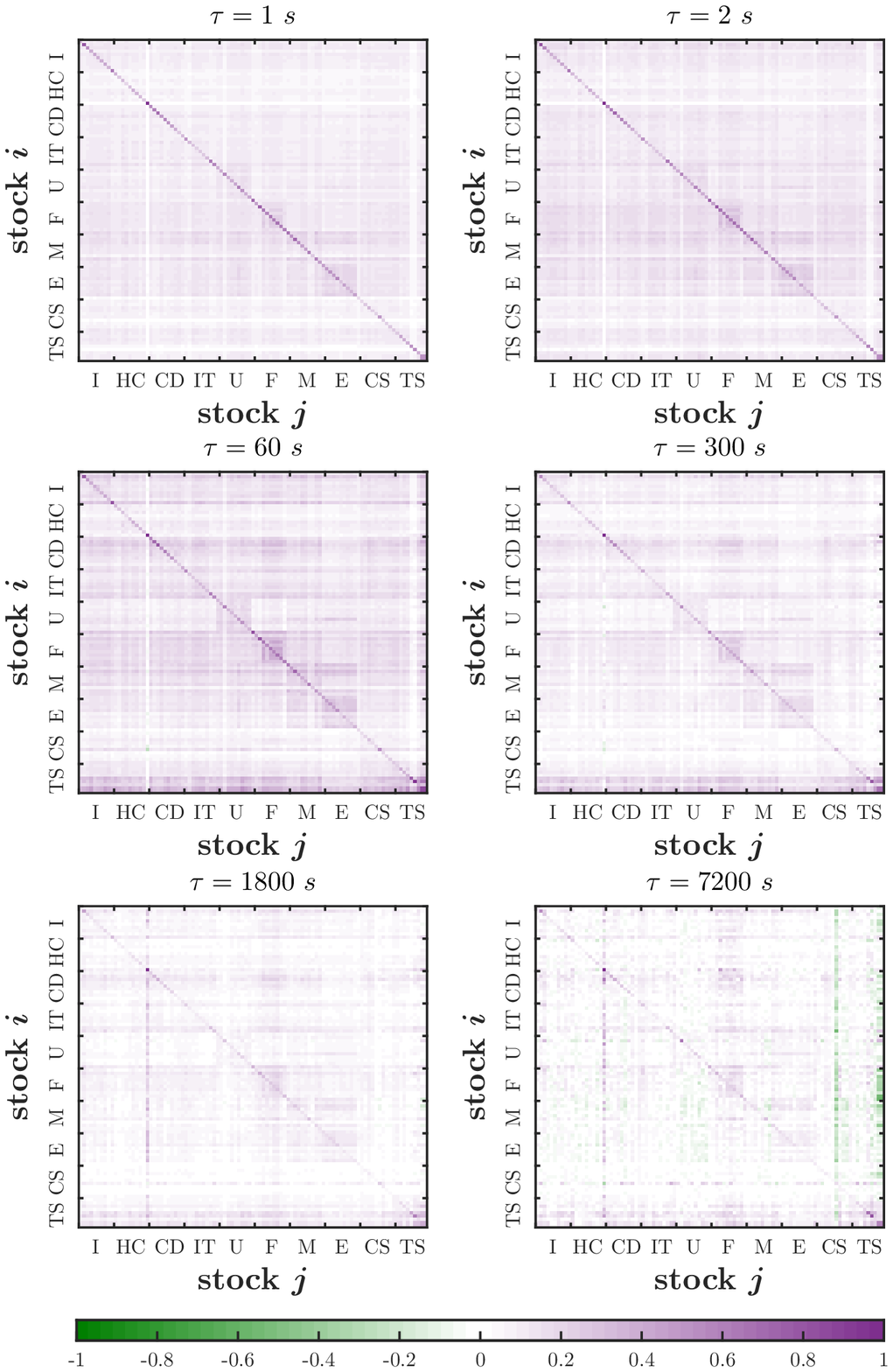}
  \end{center}
  \vspace*{-0.3cm}
  \caption{Matrices of market response with entries $\rho_{ij}(\tau)$ for $i,j=1,\ldots,99$ at different time lags $\tau=1, 2, 60, 300, 1800, 7200$~s in the year 2008. The stocks pairs $(i,j)$ belong to the sectors industrials (I), health care (HC), consumer discretionary (CD), information technology (IT), utilities (U), financials (F), materials (M), energy (E), consumer staples (CS), and telecommunications services (TS). The responses in the first two columns of panels include $\varepsilon_j(t)=0$, while in the last two columns $\varepsilon_j(t)=0$ is excluded.}
  \vspace*{-0.3cm}
 \label{fig41}
\end{figure*}

Our empirical analysis is depicted in Fig.~\ref{fig41} for a market with 99 stocks, see App.~\ref{appA}.
We show the $99\times99$ matrices of the market response for different time lags $\tau=1, 2, 60, 300, 1800, 7200$~s in the year 2008.  The diagonal strip is simply the self--response. In general, the price change of one stock is always affected by the trading of all others, and \textit{vice versa}. The stocks are ordered according to the economic sectors.

Although the responses both for $\varepsilon_j(t)=0$ in-- and excluded show similar trends of price reversion, they exhibit quite different market microstructures in each second as shown in Fig.~\ref{fig41}. This is particularly noticeable when the stocks act as the impacting stocks $j$. As some stocks yield stronger impacts than others, the matrix features striking patterns of strips with the cross--responses including $\varepsilon_j(t)=0$. They are associated with the corresponding sectors. For example, the information technology (IT) sector produces a visibly strong strip over almost all other sectors. This effect is quite stable over time. In contrast, strips cannot be found in the matrix of the cross--responses excluding $\varepsilon_j(t)=0$. The whole market displays a relatively homogeneous distribution of the response across the impacting stocks.

Furthermore, for $\varepsilon_j(t)=0$ either in-- or excluded, it is worth mentioning that the responses vary from sector to sector. For example, utilities (U), financials (F) and energy (E) respond to their own sectors with different strengths. Looking at the same sector, there is some difference as well. For example, the utilities (U) sector at $\tau=60$ has a weak response for $\varepsilon_j(t)=0$ included, but a relatively strong response for $\varepsilon_j(t)=0$ excluded. In other words, there is a price impact purely caused by the trades of the utilities (U), but due to the long non--trading time for the stocks in utilities (U), the impact is considerably weakened.

The market response structures in-- and excluding zero trade signs contain different information. The response including $\varepsilon_j(t)=0$ greatly depends on the impacting stocks $j$. In contrast, there is not an obvious difference of impacts purely caused by trades across the impacting stocks $j$ when excluding $\varepsilon_j(t)=0$. Thus, lack of trading appears as largely influencing the impact of trades when looking at the responses.

\subsection{Transient market impact}
\label{sec42}

For individual stocks, the price impact has a transient and a permanent part~\cite{Holthausen1987}. When a trade is executed, the price of an individual stock is impacted and reaches a new instantaneous price. As time evolves, the instantaneous price tends to revert to the initial price, but forming a final price which may be different from the initial one. The difference between the instantaneous price and the final price is referred to as the transient impact of a trade. The transient impact can be attributed to the price mean reversion when the order size is fixed~\cite{Bouchaud2004}. The price mean reversion can be regarded as a result of a game between liquidity takers and liquidity providers. The liquidity takers split their orders to conceal the trading information, leading to a long--memory self--correlations in the trade signs. Meanwhile, to keep their price at an appropriate position, the liquidity providers try to slowly mean revert the price. The difference between the final price and the initial price is referred to as the permanent impact of a trade. The permanent impact, which is due to the asymmetric liquidity, can be identified through a power--law between order size and price change~\cite{Lillo2004,Lillo2003,Gabaix2003,Plerou2004,Gerig2008}.
 
 \begin{figure}[tbp]
  \begin{center}
    \includegraphics[width=0.5\textwidth]{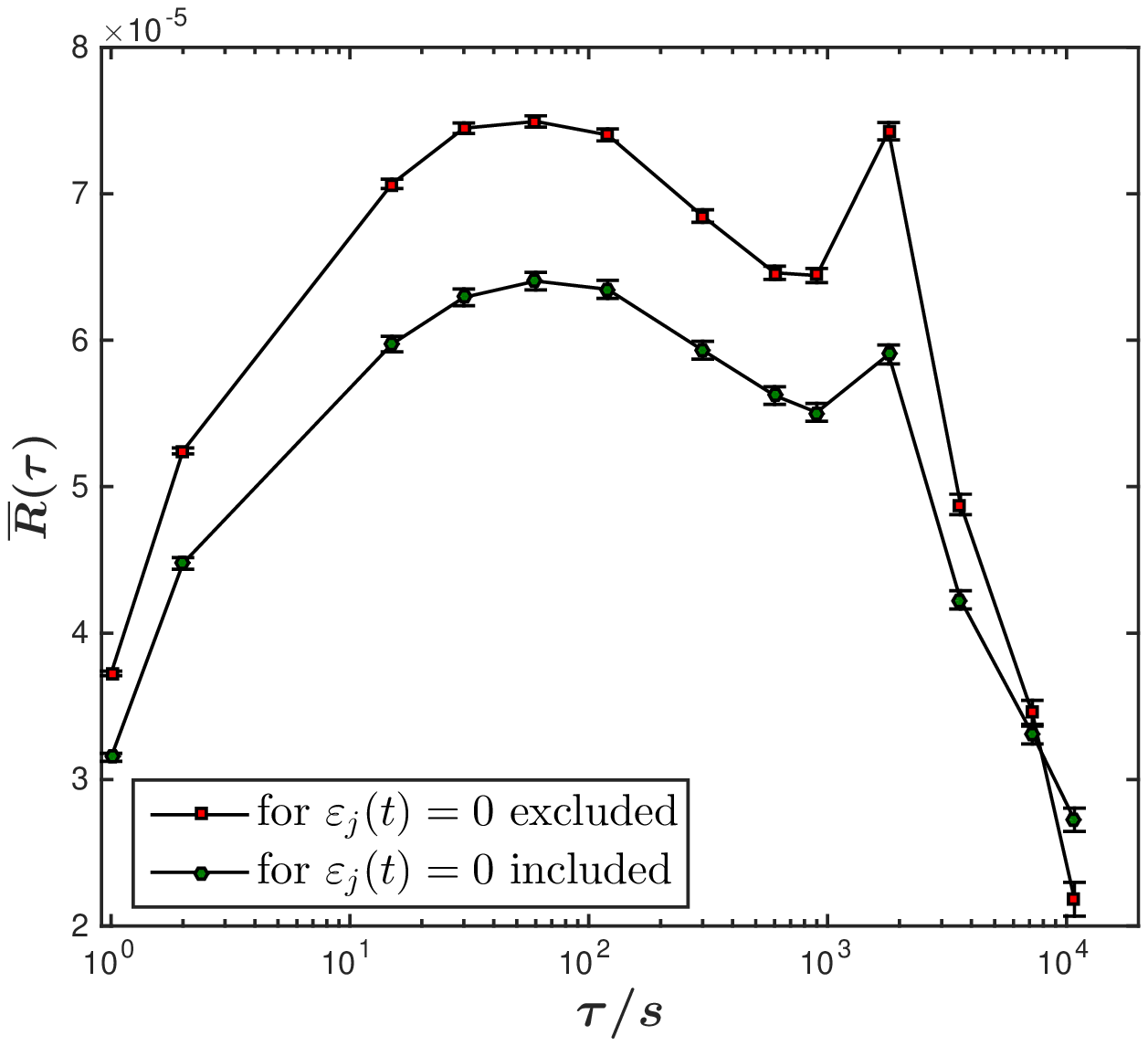}
    \vspace*{-0.3cm}
    \caption{Doubly averaged response functions $\overline{R}(\tau)$ in-- and excluding $\varepsilon_j(t)=0$ for the whole market in 2008 versus time lag $\tau$ on a logarithmic scale. The error bars indicate the standard errors. For better comparison, the doubly average response function for $\varepsilon_j(t)=0$ included and its error bars are scaled up by a factor of six.}
 \label{fig42}
  \end{center} 
    \vspace*{-0.7cm} 
\end{figure}

\begin{figure*}[tbp]
  \begin{center}
  \includegraphics[width=0.9\textwidth]{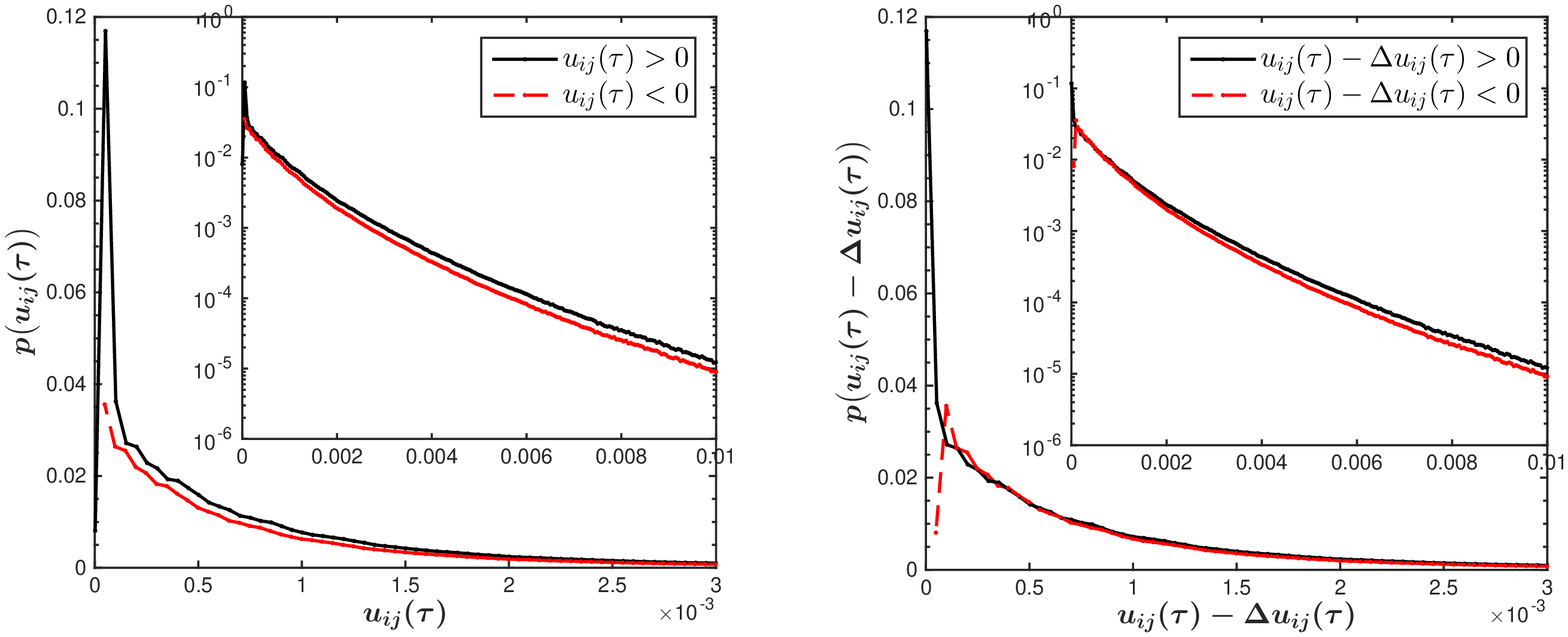}
    \caption{Probability distribution of the signed returns $u_{ij}(\tau)=r_i(t,\tau)\varepsilon_j(t)$ for the whole market in 2008, excluding $\varepsilon_j(t)=0$, for $\tau=30$ s: unshifted (left) and shifted (right) by $-\Delta u_{ij}(\tau)$ with $\Delta u_{ij}(30)=5\times10^{-5}$. The distributions for negative arguments (dashed lines) are folded back to the positive regions. A logarithmic scale is used in the insets.}
 \label{fig43}
  \end{center} 
    \vspace*{-0.5cm} 
\end{figure*}

In our study, it is difficult to identify whether a permanent impact exists or not, but the price reversion with time lag in Figs.~\ref{fig311} and \ref{fig312} as well as the evolution of the market response structure in Fig.~\ref{fig41} manifest the existence of a transient impact between stocks. As stated in Sect.~\ref{section3}, the trade of one stock cannot lower the volume in the order book and thereby change the price of another stock directly. Thus, there must be other mechanisms to explain the transient impact between stocks.

Although we cannot fully support the following statements by sufficient evidence, we wish to discuss some possible reasons for a transient impact. We thereby partly transfer a line of reasoning put forward in the case of self--responses~\cite{Toth2015} to our case of cross--responses.  One possible reason might be related to the order splitting. Suppose each stock has trade sign self--correlations, it can happen that a sequence of positive trade signs, say, in one stock partly overlaps on the physical time axis with a sequence of positive trade signs in the other stock. Such a situation can occur due to true economic correlations which are then also reflected in the return time series, or it can be purely coincidental. In any case, a cross--correlation of trade signs may result and lead to a cross--response of one stock to the trades of another stock. As the impact of trades in the self--responses is transient, the impact in cross--responses indirectly caused by sign self--correlations should be transient as well. Another possible reason might be rooted in the behavior of the traders. The overreaction to a trading information, \textit{e.g.} herding behavior, prompts traders to extend their activities to other stocks which they did not trade previously. This might also lead to price changes of those stocks. When the traders calm down and take up again their previous trading patterns, the price of those stocks will be less impacted. Here, the traders act as distributors for the trading information regarding different stocks.

\subsection{Restoring market efficiency}
\label{sec43}

Importantly, the transient impact relates to the market efficiency. According to the Efficient Market Hypothesis (EMH)~\cite{Fama1970}, the price encodes all available information, implying that arbitrage opportunities do not exist. Thus, the response functions measuring the price changes caused by trades should be zero, either for one single stock or across different stocks. However, the empirical analysis already demonstrated non--zero self--response~\cite{Bouchaud2004}. Likewise, non--zero cross--responses are seen in Fig.~\ref{fig41}. The market response is mainly positive up to time lags of about $\tau=7200$~s, while negative responses show up later. The existence of both, the empirical self-- and cross--responses, might seem to be in conflict with the EMH. However, the presence of informed trades is hardly detected for single stocks ~\cite{Bouchaud2004} and the impact of trades on the price is more likely to be due to the trading costs. Although the transient impact of trades across stocks cannot be directly traced back to the liquidity costs, we are hesitating to attribute it to a violation of the EMH. We recall that the liquidity costs, measured by the bid--ask spread, are the costs for making transactions without time delay~\cite{Bartram2008,Demsetz1968}. The trading costs include the brokerage commission and the costs resulting from the bid--ask spread, where the brokerage commission is fixed during the transaction.

Another possible explanation for the findings in Fig.~\ref{fig41} is that market efficiency is violated on short time scales, but always restored on longer time scales. As discussed in Ref.~\cite{Bouchaud2004}, the behavior of some traders will alert other traders. In this process, it is not important whether or not the trading is driven by valid information. The net effect is that the alerted traders act similar to arbitrageurs reverting the price until a state compatible with the EMH is reached. This process takes some time. Clear evidence for this interpretation is provided by the following analysis. All information is incorporated into the market as a whole, not only into the currently traded stocks, but also to the currently not traded ones. As argued in Sect.~\ref{section3}, the cross--responses fluctuate at large time lags due to a noise effect. For the whole market, these fluctuations are washed out by a self--averaging process amounting to
\begin{equation}
\overline{R}(\tau) \ = \ \langle\langle R_{ij}(\tau)\rangle _{j}\rangle _{i} \ ,
\label{eq42}
\end{equation}
where $i=j$ is excluded. Thus, the information about price changes becomes statistically much more significant. The doubly averaged response functions~\eqref{eq42} are displayed in Fig.~\ref{fig42}. The price trend is caused by a small part of potentially informed traders: first $\overline{R}(\tau)$ increases, then it decreases because of the reverting actions of the alerted traders. The decay of the average response for the whole market takes longer time than for one stock pair. This is partly a result of the noise reduction. Moreover it might indicate that the whole market needs more time to respond to all potential information than one individual stock. With time scales of about three hours, the restoration of efficiency for the whole market is a rather slow process. When the market moves back to an efficient state, the impact of trades between stocks vanishes. In this sense, the impact is transient instead of permanent both for the lack of trading in-- and excluded in the analysis.

Generalizing the analysis of Ref.~\cite{Bouchaud2004} for the individual stocks, we evaluate the evidence for the presence of informed trades across stocks by working out the probability distribution of the signed return
\begin{equation}
u_{ij}(\tau)=r_i(t,\tau)\varepsilon_j(t)
\end{equation}
for the whole market in 2008, here without $\varepsilon_j(t)=0$.  The resluts are displayed in Fig.~\ref{fig43}.  The distribution of positive and negative signed returns at $\tau=30$ s is asymmetric, but can be symmetrized by a shift of $\Delta u_{ij}(\tau)=5\times10^{-5}$. Nevertheless, for large signed returns, the asymmetry prevails.  The value of $\Delta u_{ij}(\tau)=5\times10^{-5}$ has to be compared with a value of about $\overline{R}(30) \sim 7\times10^{-5}$ for the doubly averaged response function.  As the cross--response is the average of the signed returns, the price impact can be related to at least two causes.  We find it plausible to carry over the line of reasoning for individual stocks in Ref.~\cite{Bouchaud2008} to our analysis across stocks: a shift reveals the uninformed trades while an asymmetry on longer scales hints at the presence of informed trades.

\section{Comparisons of self-- and cross--responses}  
\label{section5}
 
We now compare the self--responses and the sign self--correlators for trading and physical time scales in Sect.~\ref{sec51} as well as the self-- and cross--responses and sign cross--correlators on the physical time scale in Sect.~\ref{sec52}.

\subsection{Comparisons of self--responses on trading and physical time scales}
\label{sec51}

\begin{figure*}[htbp]
 \begin{center}
    \includegraphics[width=0.9\textwidth]{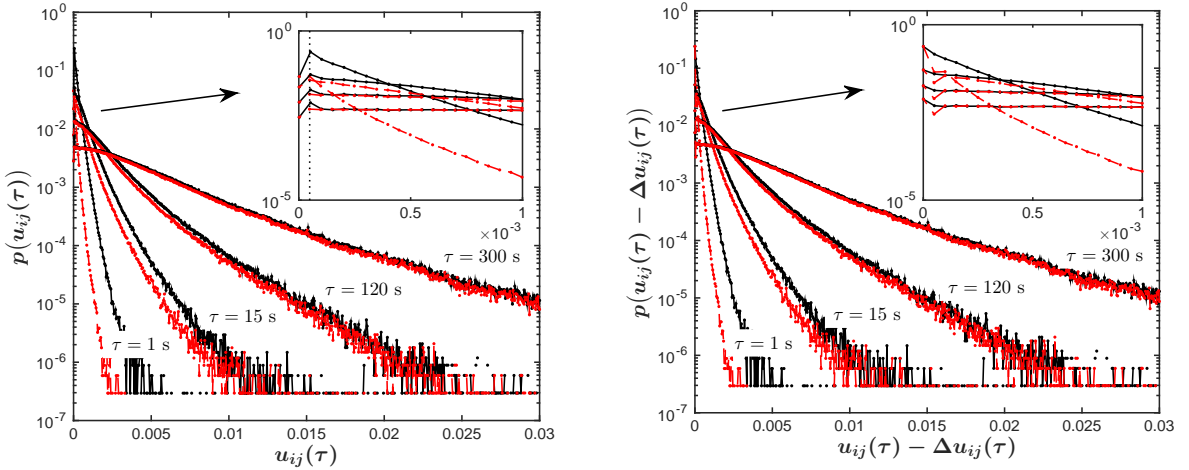}
 \end{center}
 \vspace*{-0.3cm}
 \caption{Probability distribution of signed return $u_{ij}(\tau)$ for physical time scale, where both $i$ and $j$ are AAPL in 2008: unshifted (left) and shifted by $-\Delta u_{ij}(\tau)$ (right). Here, $\Delta u_{ij}(\tau)=5.15\times10^{-5}$ with $\tau=$ 1, 15, 120, and 300 s, respectively. The negative parts of the distributions (dashed lines) are folded back to the positive regions. The insets are the amplified probability distributions near zero signed return.}
  \vspace*{-0.3cm}
\label{fig61}
\end{figure*}

As seen in Table~\ref{tab31}, the sign self--correlators on the physical time scale exhibit short memory no matter whether zero trade signs are in-- or excluded. This is very different from the sign self--correlators on the trading time scale~\cite{Bouchaud2004,Lillo2004} which have a long memory. The difference originates from the time scales used. The trading time scale regards one trade as a stamp, such that the trade sign only represents a buy or sell market order at each event stamp. The long memory of sign self--correlations caused by fragmented orders is thus easily captured even over several days. From the perspective of intraday trading, however, the trading time scale loses sight of the time between two successive trades, such that two events might have occurred within one second or separated by several hours. It is thus difficult to estimate the price impact for intraday trading. To overcome this problem, the physical time scale quantifies the price impact for a time interval of given length, \textit{e.g.} one second. The aggregated trade sign for this interval indicates an imbalance of buy and sell market orders instead of an individual buy or sell market order. Therefore, it gives limited information on the influence of multiple market orders with the same trade direction in the time interval of one second, but magnifies the effect due to only one market order with the opposite direction in another second. Hence, the long memory of sign self--correlations caused by fragmented orders is either suppressed within the one--second interval or influenced by random trades in other one--second intervals, leading to the short memory of the sign self--correlators on the physical time scale.

The value of the self--response function for $\tau=1$ s, \textit{i.e.} $R_{ii}(1)$, has to be interpreted differently on the two time scales. On the trading time scale, $R_{ii}(1)$ reflects the instantaneous impact of one trade, since a market order might not only consume the volumes in the best quotes but also the ones in the second or third best quotes to shift the price instantaneously. The price impact is not likely to come from the informed trades, it rather is due to the trading costs~\cite{Bouchaud2004}. In other words, the instantaneous impact is due to the liquidity costs on the trading time scale. However, $R_{ii}(1)$ on the physical time scale is the average impact of one second regardless of the number of trades. This impact has two causes, the uninformed trades for all lags, \textit{e.g.} liquidity costs, and the informed trades for small lags, as shown in Fig.~\ref{fig61}.

\begin{table*}[htbp]
\linespread{0.5} 
\caption{The comparisons of self-- and cross--responses} 
\begin{center}
\begin{tabular}{p{.15\textwidth}@{\hskip 0.2in}p{.2\textwidth}@{\hskip 0.2in} p{.22\textwidth}@{\hskip 0.2in} p{.22\textwidth}}
\hline
\hline	
Responses		& self--response			&self--response						& cross--response			\\
Scales			& on trading time scale		& on physical time scale				& on physical time scale		\\
\hline
\\
$\varepsilon(t)$		& the trade sign of $t$-th trade 	& the aggregated trade sign at time $t$ 	& the aggregated trade sign at time $t$ 	\\
\\
$R_{ii}(1)$ or $R_{ij}(1)$	& the instantaneous impact	& the one--second impact			& the one--second impact		\\
\\
Causes of impact	& uninformed trades	& uniformed trades and informed trades	& uniformed trades and informed trades		\\
\\
modes of impact 	& change the trade price by consuming the volumes in the order book directly	
& change the trade price both by directly consuming the volumes in the order book and indirectly affecting the placement or cancellation of limit orders	
& change the trade price by indirectly affecting the placement or cancellation of limit orders or even the execution of market orders	\\
\\
Properties of sign correlations	& long memory for sign self--correlation	& short memory for sign self--correlation	& short memory for sign cross--correlation	\\

\\
Causes of sign correlations	& main: order splitting	& main: order splitting	& probable: order splitting,	 herding behavior, portfolios	\\	 
\hline
\hline
\end{tabular}
\end{center}
\vspace*{-0.5cm} 
\label{tab62}
\end{table*}

Figure~\ref{fig61} displays the probability distribution of signed returns on the physical time scale with different time lags for AAPL in 2008, excluding the zero trade signs. The parts of the distributions for negative arguments are folded back to the positive region. There is an overall shift of the distribution from the vertical axis corresponding to zero signed returns. For the different time lags, the value of the shift are rather stable, and likely to stem from the uninformed trades. When we shift the whole distributions by $\Delta u_{ij}(\tau)=5.15\times10^{-5}$ with $\tau=$1, 15, 120, and 300 s, respectively, and then fold the negative parts of the distributions back to the positive region, we still find a difference between the two parts of the distribution, especially for those with small lags. The asymmetric tail of the distribution reflecting the imbalance of the buy and sell orders is related to the presence of the informed trades. It is only visible for small time lags, but disappears for larger ones, for example, the two parts of the distribution for $\tau=300$ s are indistinguishable. Hence the price impact in this case only stems from the uninformed trades. The informed trades can be attributed to the insufficient liquidity on short time intervals. When the liquidity is restored by submitting more limit orders, the informed trades disappear, but the remaining impact from the uninformed trades stays unchanged, indicating a possible coverage of the trading costs. Some sharp bends due to the shifting procedure can be seen for the distributions in the inset of Fig.~\ref{fig61} (right hand side) at very small signed returns.

\subsection{Comparisons of self-- and cross--responses on the physical time scale}
\label{sec52}

On the physical time scale, the trade signs at time $t$ for both, the self-- and cross--responses, are aggregated, which produce some similar features. Obviously, $R_{ii}(1)$ and $R_{ij}(1)$ both represent the impact of everything what happened in the first one second interval. Moreover, both the self-- and cross--responses contain uninformed and informed trades for all or small lags, respectively, as seen by comparing Fig.~\ref{fig43} with Fig.~\ref{fig61}.

However, the impacting stock is also impacted in the self--response, but this is not so in the cross--response, implying that there are different modes of impact. As already discussed, for the self--response, the market orders in one second move the price by removing volume of the limit orders from the order book. The therefore insufficient liquidity on short time intervals will trigger actions by the traders, affecting the placement or cancellation of limit orders or even the execution of market orders. For the cross--response, on the other hand, trades of one stock that move the price of another stock do not directly affect the volumes of the latter stock in the order book.  The impact across different stocks is likely to occur through the spread of information about the trading that influences the traders' actions. The impact due to consuming volumes is stronger than that due to the information spread, as there also is competing information, \textit{e.g.} about other trades, or relevant incoming news. Thus, the self--responses is mostly stronger than the cross--responses, as shown in Figs.~\ref{fig311} and \ref{fig312}.

Connected to the self-- and cross--responses, the sign self-- as well as cross--correlations have short memory. Regardless of the effects due to effectively random trading, the sign self--correlation can mainly be attributed to the order splitting, just like the sign self--correlation on the trading time scale. However, the sign cross--correlations may have multiple causes. One is the herding behavior. Some traders follow others when deciding to buy or sell specific stocks, but they also trade other stocks related by the same or opposite trading direction. This yields cross--correlations of trade signs across stocks. Another possible cause might be the order splitting, as discussed in Sect.~\ref{sec42}. Of course, the way how the portfolios are chosen might also be a reason, but we cannot support this by empirical evidence in our study. Table~\ref{tab62} provides a synopsis of the comparison just carried out.

\section{Conclusions}  
\label{section6}

We extended the study of stock prices responses to trading from individual stocks, \textit{i.e.} self--responses, to a whole correlated market. We empirically investigated the price responses to the trading of different stocks, \textit{i.e.} cross--responses, as functions of the time lag. The cross--response functions increase and then reverse back. Thus, the impact of the trades on the prices appears to be transient. Pictorially speaking, the market needs time to react to the distortion of efficiency caused by the potentially informed traders. In this period of distortion, some traders, who might be interpreted as arbitrageurs, drive the price to a reversion and thereby help to restore market efficiency. The cross--response is clearly related to the trade sign cross--correlations. These cross--correlators decay in a power--law fashion, revealing a short--memory process with exponents larger than one for a stock pair.

In our analysis, we preferred using the physical time instead of the trading time, as the trading of different stocks is never synchronous. We discussed the differences occurring due to the choice of time scale. We careful studied the accuracy of our trade sign classifications. In much detail, we also analyzed the effects due to including or excluding zero trade signs.

We also looked at the market as a whole by setting up a matrix, the market response, that collects the normalized information of all response functions. Several characteristic features show up which are visible in patterns having a remarkable stability in time. The market response provides quantitative information about how the trading of one stock affects the prices of other stocks, and how its own price is influenced by the trades of other stocks stocks.

Finally, we compared the self-- and the cross--responses. The self--responses on the trading and on the physical time scales show quite different characteristics. We found that the self--response on the physical scale originates from uniformed as well as from informed trades, in contrast to self--response on the trading scale that is only due to uninformed trades as already shown in previous studies. Moreover, on the physical time scale, the difference of the self-- and cross--responses is striking, especially when considering different kinds of impact.

\section*{Acknowledgements}

We thank D.~Chetalova, T.A.~Schmitt, Y.~Stepanov, and D.~Wagner for fruitful discussions. We are grateful to the referee of an earlier version of the paper for several helpful comments. One of us (S.W.) acknowledges financial support from the China Scholarship Council (grant no. 201306890014).

\section*{Author contribution statement}

T.G.~proposed the research. R.S.~and S.W.~developed the method of analysis, which S.W.~carried out. All authors contributed equally to analyzing the results, the paper was written by S.W.~and T.G.

\appendix

\section{Stocks used for analyzing the market response}
\label{appA}

We evaluated the market response for the 99 stocks from ten economic sectors: industrials (I), health care (HC), consumer discretionary (CD), information technology (IT), utilities (U), financials (F), materials (M), energy (E), consumer staples (CS), and telecommunications services (TS) as listed in Table~\ref{tabA}. The acronym AMC in Table~\ref{tabA} stands for averaged market capitalization.
\begin{table*}
\linespread{0.5}
\caption{Information of 99 stocks from ten economic sectors} 
\begin{center}
\begin{footnotesize}
\begin{tabular}{llrc@{\hskip 0.4in}llr} 
\hline
\hline
\\
\multicolumn{3}{l}{Industrials (I)} &~~& \multicolumn{3}{l}{Financials (F)}   \\
\cline{1-3}\cline{5-7}
Symbol	&Company				&AMC~ 	&	&Symbol		&Company					&AMC~	\\
\cline{1-3}\cline{5-7}
FLR		&Fluor Corp. (New)			&14414.4	&	&CME		&CME Group Inc.				&49222.9	\\
LMT		&Lockheed Martin Corp.		&12857.8	&	&GS			&Goldman Sachs Group			&21524.3	\\
FLS		&Flowserve Corporation		&12670.2	&	&ICE			&Intercontinental Exchange Inc.	&14615.3	\\
PCP		&Precision Castparts			&12447.0	&	&AVB		&AvalonBay Communities			&11081.6	\\
LLL		&L-3 Communications Holdings&12170.8	&	&BEN		&Franklin Resources				&10966.2	\\
UNP		&Union Pacific				&11920.9	&	&BXP		&Boston Properties				&10893.0 	\\
BNI		&Burlington Northern Santa Fe C &11837.5&	&SPG		&Simon Property 	Group  Inc	 	&10862.4	\\
FDX		&FedEx Corporation			&10574.7	&	&VNO		&Vornado Realty Trust			&10802.3	\\
GWW	&Grainger (W.W.) Inc.		&10416.8	&	&PSA		&Public Storage				&10147.9	\\
GD		&General Dynamics			&10035.6	&	&MTB		&M$\&$T Bank Corp.			&9920.2	\\
\cline{1-3}\cline{5-7}
\\
 \multicolumn{3}{l}{Health Care (HC)} &~& \multicolumn{3}{l}{Materials (M)}   \\
\cline{1-3}\cline{5-7}
Symbol	&Company				&AMC~ 	&	&Symbol		&Company					&AMC~	\\
\cline{1-3}\cline{5-7}
ISRG	&Intuitive Surgical Inc.		&31355.9	&	 &X			&United States Steel Corp.		&15937.7	\\
BCR		&Bard (C.R.) Inc.			&11362.7	&	 &MON		&Monsanto Co.					&14662.6	\\
BDX		&Becton  Dickinson			&10298.4	&	 &CF			&CF Industries Holdings Inc		&14075.5	\\
GENZ	&Genzyme Corp.			&9728.8	&	 &FCX		&Freeport-McMoran Cp $\&$ Gld 	&11735.7	\\
JNJ		&Johnson $\&$ Johnson		&9682.6	&	&APD		&Air Products $\&$ Chemicals		&10246.4	\\
LH		&Laboratory Corp. of America Holding&9035.7& &PX		&Praxair  Inc.					&10234.5	\\
ESRX	&Express Scripts			&8864.6	&	&VMC		&Vulcan Materials				&8700.4	\\
CELG	&Celgene Corp.			&8783.1	&	&ROH		&Rohm $\&$ Haas				&8527.1	\\
ZMH		&Zimmer Holdings			&8681.7	&	&NUE		&Nucor Corp.					&7997.4	\\
AMGN	&Amgen					&8543.0	&	&PPG		&PPG Industries				&7336.7	\\
\cline{1-3}\cline{5-7}
\\
\multicolumn{3}{l}{Consumer	Discretionary (CD)} &~& \multicolumn{3}{l}{Energy (E)}   \\
\cline{1-3}\cline{5-7}
Symbol	&Company				&AMC~ 	&	&Symbol		&Company					&AMC~	\\
\cline{1-3}\cline{5-7}
WPO	&Washington Post			&61856.1	&	&RIG		&Transocean Inc. (New)			&16409.5	\\
AZO		&AutoZone Inc.				&14463.7	&	&APA		&Apache Corp.					&13981.9	\\
SHLD	&Sears Holdings Corporation	&11759.2	&	&EOG		&EOG Resources				&13095.0	\\
WYNN	&Wynn Resorts Ltd.			&11507.9	&	&DVN		&Devon Energy Corp.			&12499.7	\\
AMZN	&Amazon Corp.			&10939.2	&	&HES		&Hess Corporation				&11990.4	\\
WHR	&Whirlpool Corp.			&9501.9	&	&XOM		&Exxon Mobil Corp.				&11460.3	\\
VFC		&V.F. Corp.				&9051.2	&	&SLB		&Schlumberger Ltd.				&11241.1	\\
APOL	&Apollo Group				&8495.8	&	&CVX		&Chevron Corp.				&11100.0	\\
NKE		&NIKE Inc.				&8149.5	&	&COP		&ConocoPhillips				&10215.3	\\
MCD		&McDonald's Corp.			&8025.6	&	&OXY		&Occidental Petroleum			&9758.4	\\
\cline{1-3}\cline{5-7}
\\
\multicolumn{3}{l}{Information Technology (IT)} &~& \multicolumn{3}{l}{Consumer Staples (CS)}   \\
\cline{1-3}\cline{5-7}
Symbol	&Company				&AMC~ 	&	&Symbol		&Company					&AMC~	\\
\cline{1-3}\cline{5-7}
GOOG	&Google Inc.				&62971.6	&	&BUD		&Anheuser-Busch				&9780.6	\\
MA		&Mastercard Inc.			&28287.8	&	&PG			&Procter $\&$ Gamble			&9711.5	\\
AAPL	&Apple Inc.				&22104.1	&	&CL			&Colgate-Palmolive				&9549.2	\\
IBM		&International Bus. Machines	&15424.9	&	&COST		&Costco Co.					&9545.9	\\
MSFT	&Microsoft Corp.			&10845.1	& 	&WMT		&Wal-Mart Stores				&9325.7	\\
CSCO	&Cisco Systems			&8731.4	&	&PEP		&PepsiCo Inc.					&9180.7	\\
INTC		&Intel Corp.				&8385.8	&	&LO			&Lorillard Inc.					&8919.0	\\
QCOM	&QUALCOMM Inc.			&7739.4	&	&UST		&UST Inc.						&8433.1	\\
CRM		&Salesforce Com Inc. 		&7691.9	&	&GIS		&General Mills					&8243.3	\\
WFR		&MEMC Electronic Materials	&7392.8	&	&KMB		&Kimberly-Clark				&8069.5	\\
\cline{1-3}\cline{5-7}
\\
\multicolumn{3}{l}{Utilities (U)} &~& \multicolumn{3}{l}{Telecommunications Services (TS)}   \\
\cline{1-3}\cline{5-7}
Symbol	&Company				&AMC~ 	&	&Symbol		&Company					&AMC~	\\
\cline{1-3}\cline{5-7}
ETR		&Entergy Corp.				&12798.7	&	&T			&AT$\&$T Inc.					&6336.2	\\
EXC		&Exelon Corp.				&9738.8	&	&VZ			&Verizon Communications		&5732.5	\\
CEG		&Constellation Energy  Group 	&9061.5	&	&EQ			&Embarq Corporation			&5318.7	\\
FE		&FirstEnergy Corp.			&8689.4	&	&AMT		&American Tower Corp.			&5195.6	\\
FPL		&FPL Group				&7742.8	&	&CTL 		&Century Telephone				&4333.8	\\
SRE		&Sempra Energy			&6940.6	&	&S			&Sprint Nextel Corp.				&2533.7	\\
STR		&Questar Corp.				&6520.4	&	&Q			&Qwest Communications  Int 		&2201.3	\\
TEG		&Integrys Energy Group Inc. 	&5978.4	&	&WIN		&Windstream Corporation			&2089.1	\\
EIX		&Edison Int'l				&5877.5	&	&FTR		&Frontier Communications		&1580.9	\\
AYE		&Allegheny Energy			&5864.9	&	&			&							&		\\
\hline
\hline
\end{tabular}
\end{footnotesize}
\end{center}
\label{tabA}
\end{table*}

\section{Error estimation}
\label{appB}
Suppose we measured or numerically simulated a set of $M$ data points $y(\tau_m)$ at positions $\tau_m, \ m=1,\ldots,M$. We want to describe the data with a function $f(\tau)$ by fitting its $M_P$ parameters. To assess the quality of the fit, the normalized
$\chi^2$~\cite{Bevington2003}
\begin{equation}
\chi^2 \ = \frac{1}{M-M_P} \sum_{m=1}^{M}\bigl(f(\tau_m)-y(\tau_m)\bigr)^2 \ ,
\label{eqB}
\end{equation}
is used. Here, $M-M_P$ is referred to as the number of degrees of freedom. In our case, we have $M=1000$, $M_P=3$ for the fitting of trade sign cross--correlators in stock pairs.

\section{Cross--response noise}
\label{appC}

\begin{figure}[tb]
  \begin{center}
  \includegraphics[width=0.49\textwidth]{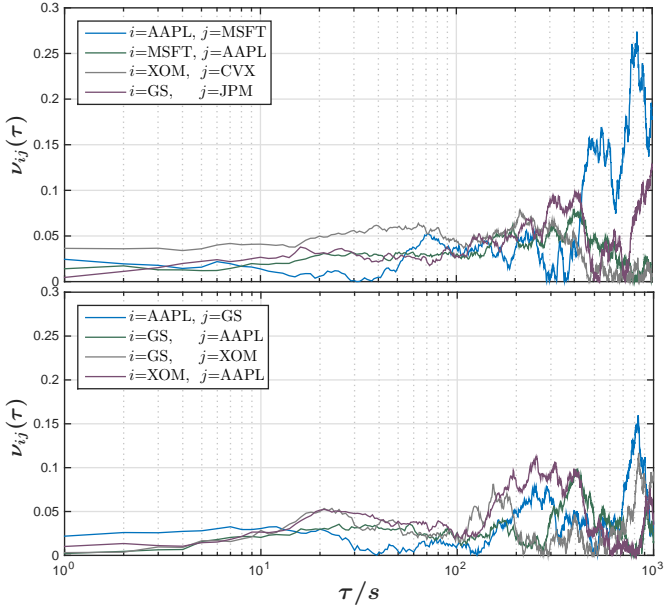}
  \end{center}
  \vspace*{-0.3cm} 
  \caption{Cross--response noise $\nu_{ij}(\tau)$ for stock different pairs during the year 2008 versus the time lag $\tau$ measured on a logarithmic scale. Stock pairs from the same economic sector (top), pairs of stocks from different sectors (bottom).}
 \label{fig34}
\end{figure}

As pointed out above, the cross--response functions and the sign cross--correlators strongly fluctuate during the decay. Here, we address this point by introducing the cross--response noise $\nu_{ij}(\tau)$ as an estimator: We determine the number $T_{ij}^{(c)}$ common to stocks $i$ and $j$ in which trading took place.  We label these days with a running integer number and separate our data into two sets, for days with even and odd numbers, respectively. We work out the corresponding cross--response functions $R_{ij}^{(k)}(\tau)$ with $k=1,2$ for the averages over odd or even days.  Each of these two functions should be very close to the
cross--response function $R_{ij}(\tau)$ averaged over all days. Thus, we introduce a cross--response noise as some kind of normalized Euclidian distance
\begin{equation}
\nu_{ij}(\tau) \ = \ \frac{1}{|R_{ij}(\tau)|}
            \sqrt{\frac{1}{2}\sum^{2}_{k=1}\left(R_{ij}^{(k)}(\tau)-R_{ij}(\tau)\right)^2} 
\label{eq37}
\end{equation}
for each value of the time lag $\tau$. In Fig.~\ref{fig34} we present the empirical results for the cross--response noise during the year 2008. Obvious, most stock pairs do not suffer from large cross--response noise for time lags smaller than about 120 seconds. During this period, the noise lies below a value of about 0.06. With increasing time lag, the noise becomes very strong, indicating unstable cross--response. The largest values the noise reaches are higher than 0.25 for lags tending towards 1000 seconds. This is the reason why some stock pairs show upwards trends after reversing back. As the sign cross--correlator weakens in the regime of large time lag, other factors dominate, leading to the large cross--response fluctuations. Limited statistics blurs the picture, since there are only 22200 seconds of effective trading time in each trading day. This clearly demonstrates that, when looking at the cross--response, the lags considered must not be too large to obtain meaningful results.

\end{document}